\documentclass[prd,aps,nofootinbib,floats]{revtex4}
\usepackage{graphicx}
\usepackage[hypertex]{hyperref}

\def\gtorder{\mathrel{\raise.3ex\hbox{$>$}\mkern-14mu
             \lower0.6ex\hbox{$\sim$}}}
\begin{document}
\title{What can we learn from neutrinoless double beta decay
  experiments?}
\author{John N. Bahcall}\thanks{jnb@ias.edu}
\author{Hitoshi~Murayama}\thanks{murayama@ias.edu}\thanks{On leave of absence
    from Department of Physics, University of California, Berkeley,
    California 94720, USA}
\author{C.~Pe\~na-Garay}\thanks{penya@ias.edu}
\affiliation{School of Natural Sciences, Institute for Advanced Study,
  Princeton, NJ 08540}
\date{March 13, 2004}
\begin{abstract}
  We assess how well next generation neutrinoless double beta decay
  and normal neutrino beta decay experiments can answer four
  fundamental questions. 1) If neutrinoless double beta decay searches
  do not detect a signal, and if the spectrum is known to be inverted
  hierarchy, can we conclude that neutrinos are Dirac particles? 2) If
  neutrinoless double beta decay searches are negative and a next
  generation ordinary beta decay experiment detects the neutrino mass
  scale, can we conclude that neutrinos are Dirac particles? 3) If
  neutrinoless double beta decay is observed with a large neutrino
  mass element, what is the total mass in neutrinos? 4) If
  neutrinoless double beta decay is observed but next generation beta
  decay searches for a neutrino mass only set a mass upper limit, can
  we establish whether the mass hierarchy is normal or inverted? We
  base our answers on the expected performance of next generation
  neutrinoless double beta decay experiments and on simulations of the
  accuracy of calculations of nuclear matrix elements.
\end{abstract}
\maketitle

\renewcommand{\baselinestretch}{1.2}
\section{Introduction}
\label{sec:introduction}

A new generation of double beta decay experiments will be
undertaken with unprecedented accuracy. In approximately the same
time frame,  it will become possible to make much more precise
measurements of, or set constraints on, the mass of neutrinos
emitted in ordinary beta decay. The results of these next
generation experiments will be important for understanding the
physics of weak interactions.

If neutrinoless double beta decay is observed, then one can
conclude~\cite{majoranacharacter} immediately that neutrinos are
Majorana particles without messing around with detailed calculations
and qualifications of the kind discussed in this paper. (We will not
consider alternative interpretations such as $R$-parity violation
\cite{Hall:1983id,Mohapatra:su,Mohapatra:1981pm},
which can probably be verified or excluded at
high-energy colliders.  The violation of lepton number is
clear in either case.)  The community of physicists can and will
celebrate if double beta decay is observed.

In this paper, we provide quantitative estimates of how well we
can answer four other fundamental questions about neutrinos using
the assumed results of the next generation of neutrinoless double
beta decay experiments and normal beta decay experiments.

Our principal results are summarized in Table~\ref{tab:summary}.

\begin{table}[t]
\caption{ Answers to some questions about the potential of
neutrinoless double beta decay experiments. Answers refer to a CL
of 99.73 \%CL for the assumed probability distributions. We adopt
a sensitivity $s$ equal to what is projected for the Majorana
experiment \protect\cite{majorana} (if another reference
sensitivity $s'$ is assumed, the required number of experiments
should be scaled by $N'_{exp}=N_{exp} s/s'$). If the answer for an
inverted neutrino mass hierarchy is different from the answer for
a normal mass hierarchy (see Fig.~\ref{fig:hierarchy}), we show in
parentheses the answer for a normal mass hierarchy.
\label{tab:summary}}
\begin{tabular}{cccc}
Section & Assumptions & Question  &$N_{\rm exp}$ at 99.73 \%CL\\
\noalign{\smallskip}\hline
\noalign{\medskip} II & No  detected neutrinoless double $\beta$-decay  &  Dirac ? & 230 $(\infty$) \\
\noalign{\smallskip}\hline
\noalign{\medskip} III & lightest mass scale (1 $\pm$ 0.05 eV), ~No  neutrinoless double $\beta$-decay &  Dirac ? &  1 \\
\noalign{\medskip} III & lightest mass scale (0.35 $\pm$ 0.07 eV), ~No  neutrinoless double $\beta$-decay  &  Dirac ? &  5 (6) \\
\noalign{\medskip} III & lightest mass scale (0.3 $\pm$ 0.1 eV), ~No  neutrinoless double $\beta$-decay &  Dirac ? & 16 ($\infty$) \\
\noalign{\smallskip}\hline \noalign{\medskip} IV &  Neutrinoless
double $\beta$-decay: ~$T_{\frac{1}{2}} (^{76}{\rm Ge})= (3.2 \pm
0.2) \times 10^{25}$ yr
& Total mass ? &  [0.46,9.56] ([0.48,9.58])\\
\noalign{\medskip} IV & Neutrinoless double $\beta$-decay:
~$T_{\frac{1}{2}} (^{76}{\rm Ge})= (1.\pm 0.1) \times 10^{26}$ yr
& Total mass ? &  [0.24,8.34] ([0.28,8.40])\\
\noalign{\medskip} IV &  Neutrinoless double $\beta$-decay:
~$T_{\frac{1}{2}} (^{76}{\rm Ge})= (3.2 \pm 0.5) \times 10^{26}$
yr
& Total mass ? &  [0.08,5.68] ([0.16,6.06])\\
\noalign{\smallskip}\hline \noalign{\medskip} V & Detected
neutrinoless double $\beta$-decay
&  Hierarchy ? &  No \\
\noalign{\medskip} V & Detected neutrinoless double $\beta$-decay,
~{\rm ~ private~communication:}~ m~=~0&  Hierarchy ? &  Yes \\
\end{tabular}
\end{table}

\subsection{How can we estimate the uncertainties in calculated nuclear
matrix elements?} \label{subsec:introuncertainties}

The uncertainty in the calculated nuclear matrix elements for
neutrinoless double beta decay will constitute the principal
obstacle to answering some basic questions about neutrinos.  The
essential problem is that the correct theory of nuclei is QCD, a
notoriously difficult theory to do calculations with for nuclei
with several nucleons. For neutrinoless double beta decay, the
situation is even more severe since double beta candidates involve
systems with $A \sim 50$ to $A \sim 100$ and even larger. Very
attractive next generation experiments have been proposed for a
number of different isotopes, including $^{48}$Ca~\cite{KIS01},
$^{76}$Ge~\cite{majorana,ZDE01,KLA01b},
$^{100}$Mo~\cite{NEMO3,EJI00}, $^{116}$Cd~\cite{BEL01},
$^{130}$Te~\cite{ZUB01,AVI}, $^{136}$Xe\cite{XMASS,CAC01,Gratta},
$^{150}$Nd~\cite{ISH00}, and $^{160}$Gd~\cite{DAN01,WANGS01}.

In the foreseeable future, it does not seem possible to derive in
a  direct and controlled manner from QCD nuclear matrix elements
for large $A$. Thus there is no way of quantifying with absolute
confidence the range of uncertainties in nuclear matrix elements
calculated with different theoretical models or approximations.

In the absence of being able to derive the errors directly from
QCD, we assume that the published range of calculated matrix
elements defines a plausible approximation to the uncertainty in
our knowledge of the matrix elements. We do not, for example,
favor a particular calculation because it happens to give better
agreement with the inferred matrix element for two-neutrino double
beta-decay (in the rare cases where this decay has been observed).
We have no way of knowing for sure what the improved agreement for
the two-neutrino case implies for the neutrinoless double-beta
decay matrix element and whether, indeed, the agreement in a
special case is accidental or not\footnote{Fukugita and
Yanagita~\cite{fukugitabook} note that the nuclear levels that are
important for neutrinoless double beta-decay are typically at
excitation energies of order 10 MeV, while for two neutrino double
beta-decay the characteristic excitation energies are lower, a few
MeV. Thus even if the lower excitation states are correctly
described, there is no guarantee that the higher excitation states
are also correctly described.}.

We recognize that different individuals may regard the calculated
range of nuclear matrix elements as either too narrow or too broad
to reflect the actual uncertainty. However, we do not know of any
way to settle objectively and conclusively whether our estimate of
the uncertainty is pessimistic or optimistic in any particular
case.

\subsection{Some definitions}
\label{subsec:somedefinitions}

The neutrino mass matrix element that appears in neutrinoless
double beta decay \cite{Vogel,ev,rev} is given by
\begin{equation}
|\langle m^\nu_{ee} \rangle| ~=~
m_e\frac{1}{\sqrt{T_{\frac{1}{2}}F_N}} ~=~m_e
\sqrt{{\frac{\lambda}{\ln2 F_N}}} \, ,
\label{eq:matrixelementdefn}
\end{equation}
where $m_e$ is the electron mass, $T_{\frac{1}{2}} (\lambda)$ is
the half life (exponential decay constant) of the double beta
decay process, and the nuclear structure parameter $F_N$ is given
by

\begin{equation}
F_N =  G^{0\nu} \left|M_F^{0\nu} - \left(\frac{g_A}{g_V}\right)^2
M_{GT}^{0\nu}\right|^2 \, .
\end{equation}

For specificity, we consider a neutrinoless double beta decay
experiment with sensitivity to $T_{\frac{1}{2}}F_N$ that is
exemplified by what is expected for the Majorana experiment
\cite{majorana} (see also, compilation in Ref.~\cite{ev}).
We will consider that a number $N_{exp}$ of neutrinoless double beta
decay experiments are performed with the expected Majorana
sensitivity $s$.

Our results are, however, general. If the experiments that
actually are or could be performed have a different sensitivity,
then our results should be rescaled by

\begin{equation}
N'_{exp}~=~ N_{exp} s/s'\, . \label{eq:rescale}
\end{equation}
If a specific neutrinoless double beta-decay experiment
successfully detects a signal, then a greatly increased exposure
with the same detector will not improve much the confidence with
which one can answer the questions raised in this paper. For a
single detector, the uncertainty will be dominated by the nuclear
factor of that nucleus. Measurements with different nuclei will be
required to improve the statistical significance of the answers to
questions about the nature and properties of neutrinos. On the
other hand, suppose the search for neutrinoless double beta decay
is negative with a given detector. Then an increase in the
exposure time by a factor $N_{\rm exposure}$ is equivalent to
performing $N_{\rm exposure}$ new experiment that have the
identical sensitivity.

\begin{figure}[!t]
  \centerline{\includegraphics[width=3.75in]{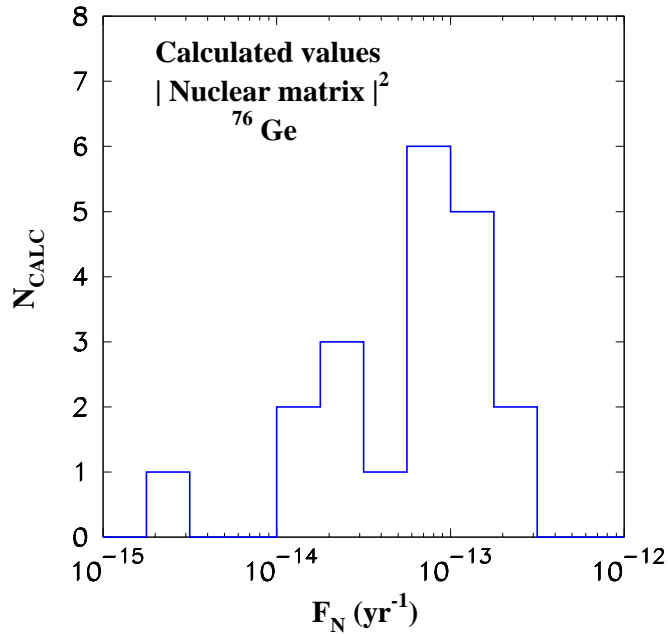}}
  \caption{Distribution of $^{76}$Ge nuclear factor results. A
    compilation of 20 different calculations spans in the range $2.7
    \times 10^{-15} - 2.9 \times 10^{-13}$ yr$^{-1}$
    \protect\cite{majorana} }. \label{fig:fn}
\end{figure}

The neutrino mass element $|\langle m^\nu_{ee} \rangle|$ is
related to the fundamental neutrino parameters by the expression
\begin{equation}
|\langle m^\nu_{ee} \rangle| ~=~ \left| m_1 |U^2_{e1}| e^{i\phi_1} +
m_2 |U^2_{e2}| e^{i\phi_2} +  m_3 |U^2_{e3}| \right| \,
,\label{eq:massIH}
\end{equation}
where $m_i$ are the mass eigenvalues of the Majorana neutrinos, $U$
is the lepton mixing matrix, and $\phi_i$ are relative Majorana
phases. Normal (Inverted) hierarchy corresponds to
 the ratio between mass eigenvalues (labelled in increasing mass
eigenvalue $m_1<m_2<m_3$) given by $m_3/m_2>m_2/m_1$ ($m_3/m_2<m_2/m_1$).
If hierarchies are indistinguishable, what happens when
$\Delta m^2_{ij} << m^2_1$, then the mass scheme is called degenerate.

\subsection{The dispersion in calculated nuclear matrix elements}
\label{subsec:disperioncalculated}

The dispersion of the calculated nuclear matrix elements obtained
by different theoretical methods is large. For example,  a
compilation of 20 different calculations
\cite{majorana,Tretyak:1995rj,Civitarese:2002tu} for
$^{76}$Ge spans the range $2.7 \times 10^{-15} - 2.9 \times
10^{-13}$ yr$^{-1}$.

Figure~\ref{fig:fn} shows the distribution of $^{76}$Ge nuclear
factors binned in a logarithmic scale. In our analyses of how much
we can learn about different fundamental neutrino questions, we
will also consider $F_N$ as a random
 variable in linear and logarithmic scales of the constant and the Gaussian
probability distributions. For the Gaussian distribution, we will
adopt  the central value of the $F_N$ interval as the mean and one
third of the radius of the interval covered by calculated values
of $F_N$ as the standard deviation. The lowest nuclear factor
$F_N$ shown in Fig.~\ref{fig:fn} corresponds to a recent
calculation~\cite{Bobyk:2000dw} that used a self-consistent
renormalized quasi-particle random phase approximation. We do not
know of any rigorous argument that would exclude this recent
calculation while including the other calculations shown in the
figure.

 For the numerical
calculations given in this paper,  we used the distribution
 of calculated nuclear factor for  $^{76}$Ge because this nucleus is the one
for which we found the largest number of published calculations of
$F_N$. We performed Kolmogorov-Smirnov tests to test if the
distributions of $F_N$ that were calculated for other double
beta-decay candidates ( $^{82}$Se, $^{130}$Te, $^{136}$Xe) are
consistent with the distribution shown in Fig.~\ref{fig:fn}
 of $^{76}$Ge. Table 2 of Ref.~\cite{ev} compiles a list of six calculations
\cite{calc} for these nuclei. The Kolmogorov-Smirnov tests show
that we can not
 reject at 95\%CL, for any of the nuclei $^{82}$Se, $^{130}$Te, or
 $^{136}$Xe,
 the hypothesis that the distribution of
calculations of $F_N$ given in Table 2 of Ref.~\cite{ev} is the
same distribution as shown in Fig.~\ref{fig:fn} for $^{76}$Ge.
We also checked that the distribution of the six calculations
listed in Table 2 of Ref.~\cite{ev} for $^{76}$Ge is consistent
with the distribution of  20 calculations of $F_N$
 used  in the present work.

The fact that the uncertainty in the nuclear matrix element plays
a major role in our ability to resolve fundamental questions in
neutrinoless double beta decay experiments is well known (see for
example the famous reviews in Ref.~\cite{rev}).
Reference~\cite{Bilenky:2004wn} is the most recent example with
which we are familiar of a systematic analysis that assumes a
small uncertainty in the nuclear matrix elements for neutrinoless
double beta-decay experiments (for the nuclear physics discussion
see Ref.~\cite{Rodin:2003eb}). The discussion in
Ref.~\cite{Bilenky:2004wn} assumes the correctness of the
renormalized quasi-particle random phase approximation (RQRPA)
that leads to the lowest nuclear factor in Fig.~\ref{fig:fn}.
Readers who are optimistic regarding the validity of current
calculational methods for calculating nuclear matrix elements in
neutrinoless double beta-decay may prefer the conclusions of
Ref.~\cite{Bilenky:2004wn} instead of the more conservative
conclusions of the present paper.

 The position adopted in this
paper is that the RQRPA could be accurate, or some other
calculational scheme could be more accurate, but we will not know
for sure how precise any approximation is until calculations can
be done in a controlled manner using QCD.  Our attitude is
consistent with the point of view expressed in the recent
discussion of the RQRPA and QRPA approximations in
Ref.~\cite{Rodin:2003eb}.  These authors summarized their analysis
with the statement~\cite{Rodin:2003eb}: ``Even though we cannot
guarantee this basic method [RQRPA] is trustworthy, we have
eliminated, or at least greatly reduced, the arbitrariness
commonly present in published calculations.'' In other words, the
recommended prescription results in a small dispersion in
calculated nuclear matrix elements, which may or may not be close
to the true value.

The reader will chose what to believe based upon the reader's
convictions about the accuracy of the calculations of nuclear
matrix elements. We believe that the burden of proof is upon the
person drawing conclusions that depend upon the size of the
nuclear matrix elements. The conclusions must be supported by a
proof that the matrix elements are equal to the QCD values within
the stated errors.

\begin{figure}[!t]
\begin{center}
  \includegraphics[width=3.75in]{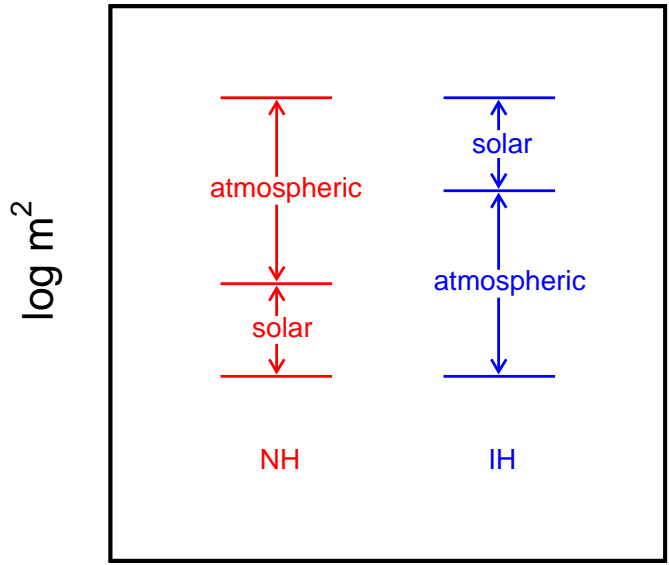}
\end{center}
  \caption{\label{fig:hierarchy} Normal and inverted neutrino mass hierarchy.
  The larger splitting is $\Delta m^2_{\rm atm} \simeq 2
    \times 10^{-3}$~eV$^2$;  the smaller splitting is $\Delta m^2_{\rm
      solar} \simeq 7 \times 10^{-5}$~eV$^2$. The hierarchy is referred as "Degenerate"
      if the square of the smallest mass is much larger than either $\Delta m^2_{\rm atm}$
      or $\Delta m^2_{\rm
      solar}$.}
\end{figure}

Our goal is to provide, for the reader's consideration, an
alternative viewpoint to the one that is usually adopted in
discussing neutrinoless double-beta decay experiments. As far as
we know, there is no previous systematic, quantitative study to
evaluate the impact of the uncertainty in the nuclear matrix
element  for different assumed probability distributions.
Recently, it has been demonstrated that it is not practical to
 detect in neutrinoless double beta-decay experiments neutrino CP
violation arising from Majorana phases \cite{barger,Pascoli:2002qm}.

\subsection{How do we determine how many experiments are
required?} \label{subsec:howmany}

For each question about neutrino properties that we address, we
make specific assumptions about what is, or is not, observed
experimentally. Depending upon the particular question we are
addressing, we will assume that the neutrino masses satisfy a
normal or an inverted hierarchy, as illustrated in
Fig.~\ref{fig:hierarchy}. We will also make assumptions regarding
the observation, or non-observation, of a neutrino mass in
ordinary (tritium) beta-decay.

\begin{figure}[!t]
  \centerline{\includegraphics[width=3.75in]{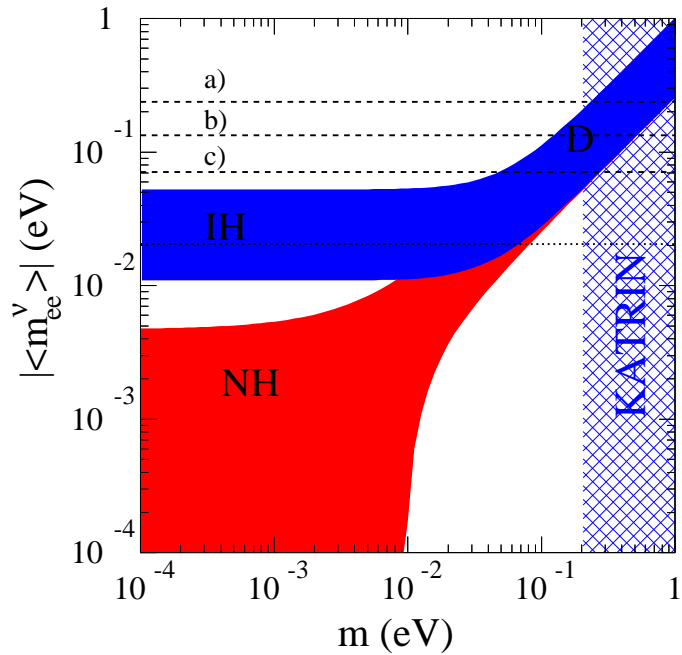}}
  \caption{The neutrinoless double beta decay mass element $|\langle m^\nu_{ee}
    \rangle|$ versus the lowest neutrino mass $m$. The regions allowed
    at 90\% CL by existing neutrino oscillation data are shown for a
    normal neutrino hierarchy (NH), an inverted hierarchy (IH), and
    degenerate neutrinos (D) (see Fig.~\ref{fig:hierarchy} for an
    explanation of the different hierarchical arrangements of neutrino
    masses and Sec.~\ref{sec:question4} for a description of how the
    allowed regions were computed). The hatched area shows the
    parameter space that can be excluded by the Katrin
    experiment\protect \cite{KATRIN} if no evidence for a neutrino
    mass is detected in tritium beta decay. The three dashed lines
    labelled a), b), and c) refer to three possible positive results
    for a next generation neutrinoless double beta decay search and
    are discussed in Sec.~\ref{sec:question3} and
    Sec.~\ref{sec:question4}. The dotted horizontal line near
    $10^{-2}$ eV illustrates the sensitivity that is expected for the
    Majorana experiment \protect \cite{majorana}. For an original version
    of this figure, see Ref.~\cite{recent2}.} \label{fig:mee_m1}
\end{figure}

Figure~\ref{fig:mee_m1} shows the relationship between the
neutrinoless double beta decay mass element $|\langle m^\nu_{ee}
\rangle|$ and the smallest neutrino mass $m$ \cite{recent2,recent3,recent4}.
This figure plays a
key role in our discussion; we will return to
Fig.~\ref{fig:mee_m1} in Sec.~\ref{sec:question1},
Sec.~\ref{sec:question3}, and Sec.~\ref{sec:question4}.

For a given set of assumptions as described above, we compute the
different probability distributions that are implied by the
assumed experimental constraints. In the final step of our
analysis, we combine the computed probability distributions in
order to determine how many experiments are required to answer a
stated question at a specific confidence level.

\subsection{What is the bottom line?}
\label{subsec:bottomline}

Some readers will only care about the bottom line. How many
neutrinoless double beta-decay experiments are required in order to
determine whether neutrinos are Majorana or Dirac
particles?\footnote{Note that we are referring to the dominant
  neutrino masses relevant to the currently observed neutrino
  oscillation.  Even if the dominant masses are Dirac, there may be
  much smaller Majorana masses not relevant to neutrino oscillation,
  sometimes called pseudo-Dirac.  We do not distinguish Dirac and
  pseudo-Dirac neutrinos in this paper.}  What fraction of the closure
mass of the universe do neutrinos constitute? Can we establish whether
the neutrino masses satisfy a normal or an inverted mass hierarchy?

Table~\ref{tab:summary} summarizes our numerical results.

We state in column 2 of Table~\ref{tab:summary} the different
assumptions that we have made about future experiments.  In column
3, we give abbreviated names to the questions that we have asked.
Finally, in column 4, we present a brief summary of our answers to
the different physical questions about neutrinos. The reader
interested in the details of how a specific question was answered
can look in the section of this paper that is listed in column 1
of Table~\ref{tab:summary}.

\subsection{Outline of this paper} \label{subsec:outline}

In Sec.~\ref{sec:question1}, we show that an impractically large
number of neutrinoless double beta decay experiments would be
required to show that neutrinos are Dirac particles if next
generation experiments do not reveal neutrinoless double beta
decay. We show in Sec.~\ref{sec:question2} that non-observation of
neutrinoless double beta decay taken together with a measurement
in ordinary beta decay of the lowest neutrino mass that is near
the present upper limit (e.g., $\sim 1$ eV) would be sufficient to
show that neutrinos are Dirac particles. However, if the neutrino
mass is as low as $0.3$ eV or lower, then many neutrinoless double
beta decay experiments would be required to show that neutrinos
are Dirac particles. We present in Sec.~\ref{sec:question3} the
allowed ranges in the total mass in neutrinos if neutrinoless
double beta decay is detected at different possible half-lives.
Finally, we show in Sec.~\ref{sec:question4} that even if
neutrinoless double beta decay is observed in next generation
experiments we nevertheless will not be able to decide from beta
decay experiments alone whether the mass hierarchy is normal or
inverted.  We discuss our principal results in
Sec.~\ref{sec:summary}.

\section{Are neutrinos Dirac particles? No neutrinoless double beta
decay and inverted hierarchy}
\label{sec:question1}

In this section,  we assume that next generation
experiments~\cite{Gratta} will not observe neutrinoless double
beta decay. Figure~\ref{fig:mee_m1} shows that it is much easier
to observe neutrinoless double beta decay if the neutrino mass
hierarchy is inverted. If the hierarchy is normal, then the
neutrino mass matrix element, $|\langle m^\nu_{ee} \rangle|$, can
be unobservably small even if neutrinos are Majorana particles,
making it impossible to decide for a normal hierarchy whether
neutrinos are Dirac are Majorana. Hence, we concentrate our
numerical calculations in this section on the case in which the
mass hierarchy is known to be inverted from long baseline
experiments~\cite{LBL,futureLBL} or from some other measurement.

\begin{figure}[!t]
  \centerline{\includegraphics[width=4.5in]{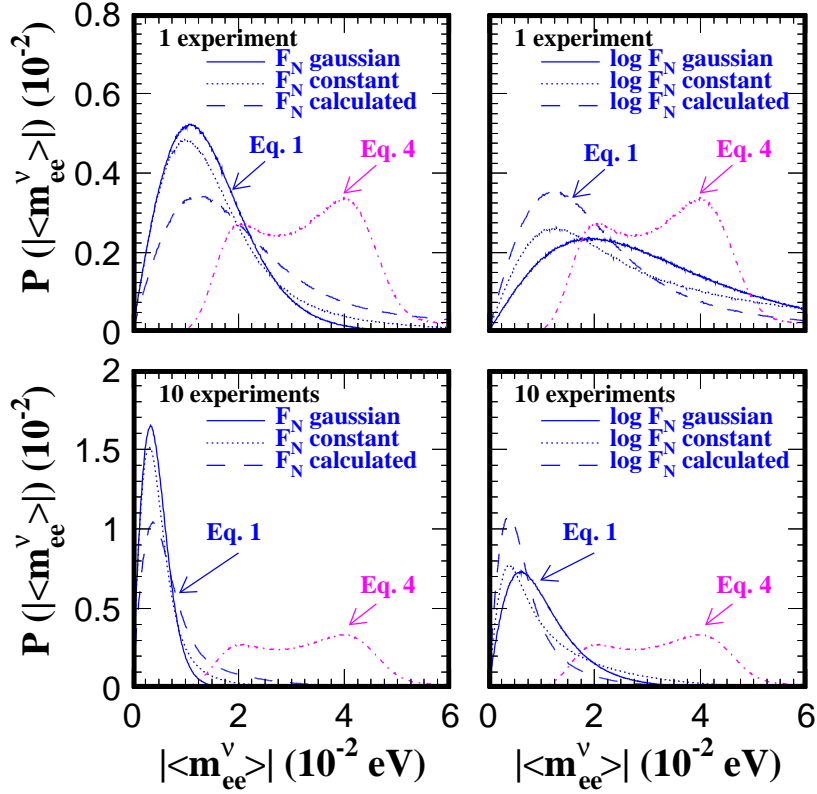}}
  \caption{Probability distributions of $|\langle m^\nu_{ee}
    \rangle|$ in future generation neutrinoless double beta decay
    experiments. On the right side of each of the four panels, we plot
    the pdf of $|\langle m^\nu_{ee} \rangle|$ that is obtained from
    neutrino oscillation data by using Eq.~(\ref{eq_mee_ih})
    (dashed-dotted line). On the left side of each of the four panels
    of the figure, we show probability distribution functions for
    $|\langle m^\nu_{ee} \rangle|$ assuming that next generation
    experiments do not detect neutrinoless double beta decay. The
    plotted pdfs were obtained by making different assumptions
    regarding the pdf of the nuclear factor $F_N$ that appears in
    Eq.~(\ref{eq:matrixelementdefn}).  For the left hand panels of
    Fig.~\ref{fig:question1}, we assume that $F_N$ follows a Gaussian
    distribution (full line), a constant distribution (dotted line),
    or the actual computed distribution of values of $F_N$ computed by
    different nuclear theorists (dashed line).  For the right hand
    panels, we assumed that $\log F_N$ follows these same
    distributions.  The upper pair of panels corresponds to a single
    experiment with a sensitivity equal to what is expected for the
    Majorana experiment \protect\cite{majorana}. The lower panels
    correspond to simulated results for ten experiments each with a
    sensitivity equal to the anticipated sensitivity of the Majorana
    experiment \protect\cite{majorana}.} \label{fig:question1}
\end{figure}

For definiteness and in order to minimize the number of required
experiments, we assume that the data are free of all background
and that there are no candidate neutrinoless double beta-decay
events. The decay constant $\lambda$ then satisfies an exponential
probability decay function (pdf) corresponding to the Poisson
probability that no events are observed.

Given that we know that there is an inverted mass hierarchy for
neutrinos, how many neutrinoless double beta decay experiments
would be needed to establish that neutrinos are Dirac particles at
a given CL? We shall see that in this case 230 neutrinoless double
beta-decay experiments are required in order to establish that
neutrinos are Dirac particles at a CL equivalent to $3\sigma$. If
we admitted that there is a possibility that the neutrino
hierarchy is normal, then an essentially infinite number of
experiments would be required.

In order to calculate the required number of experiments, we first
compute the probability distribution function of the neutrino mass
element $|\langle m^\nu_{ee} \rangle|$ given by
Eq.~(\ref{eq:matrixelementdefn}). This pdf depends upon the assumed
distribution of the nuclear matrix element $F_N$.

Figure~\ref{fig:question1} illustrates the dependence of the
probability distribution function of $|\langle m^\nu_{ee}
\rangle|$ on the pdf of the nuclear factor $F_N$. We show the
calculated pdfs for $|\langle m^\nu_{ee}\rangle|$ that follow from
Eq.~(\ref{eq:matrixelementdefn}) for different assumptions about the
pdf of $F_N$: a Gaussian (full line), a constant probability
spanning the entire range of calculated $F_N$ (dashed line), and a
pdf equal to the actual reported distribution of $F_N$(dashed
line). In the computations shown in the upper and lower left hand
corners of Fig.~\ref{fig:question1}, the value of $F_N$ is treated
as the random variable with the illustrated pdf, while in the
upper and lower right hand corners of Fig.~\ref{fig:question1},
the pdfs were calculated treating the logarithm of $F_N$ as the
random variable.  The two lower panels of Fig.~\ref{fig:question1}
are similar to the two upper panels except for the fact that the
lower panels refer to the pdfs computed assuming 10 equivalent
experiments (equal sensitivity) have been performed instead of
just one experiment.

We next concentrate on the neutrino mass element $|\langle
m^\nu_{ee} \rangle|$ as a
 function of the neutrino parameters. In the case of inverted hierarchy,
the appropriate expression for $|\langle m^\nu_{ee} \rangle|$ is
given by :

\begin{equation}
|\langle m^\nu_{ee} \rangle|_{IH} = \left| m \sin^2\theta_{13} +
\cos^2\theta_{13} ( \cos^2\theta_\odot \sqrt{m^2+\Delta m^2_{atm}
-\Delta m^2_\odot} ~e^{i\phi_1} + \sin^2\theta_{\odot}
\sqrt{m^2+\Delta m^2_{atm}} ~e^{i\phi_2})\right| \, ,\label{eq_mee_ih}
\end{equation}
where $m$ is the mass of the lowest mass eigenstate and $\Delta
m^2_\odot$ and $\Delta m^2_{atm}$ are mass squared splittings, and
$\theta_\odot$ and $\theta_{13}$ are mixing angles determined by
solar, atmospheric, reactor and K2K experiments \cite{ggpg}. We
have computed numerically the pdf of the neutrino mass element
that corresponds to  Eq.~(\ref{eq_mee_ih}). In this computation, we
used gaussian distributions for the mass squared splittings and
mixing angles, with mean values and standard deviations given by
$\Delta m^2_\odot = (7.1 \pm 0.7) \times 10^{-5}$ eV$^2$, $\Delta
m^2_{atm}  = (2.0 \pm 0.4) \times 10^{-3}$ eV$^2$, $\sin^2
\theta_\odot = 0.30 \pm 0.03$, and $\sin^2 \theta_{13} = 0.008 \pm
0.02$ \cite{ggpg,mpg,roadmap}.
 In the latter case, we truncate the gaussian distribution to include only positive values.
 We assumed constant probability distributions for the lightest mass
 $m$ (in logarithmic scale,
with $10^{-6} < m < 2.3$ eV ) and the phases $\phi_1$ and $\phi_2$
(in linear scale).

In order to compute the number of experiments required to
establish that neutrinos are Dirac particles, we must compute the
the joint probability $P$ that follows from the unobserved
neutrinoless double beta decay experiments [see
Eq.~(\ref{eq:matrixelementdefn})] and from the expression for the
neutrino mass $|\langle m^\nu_{ee} \rangle|$ in terms of various
neutrino oscillation parameters [see Eq.~(\ref{eq:massIH})]. The
probability $P$ that these two conditions are satisfied is
 given by the product of the probabilities of the two
individual constraints, conveniently normalized. Thus
\begin{equation}
P (N_{exp})= \frac{\int d|\langle m^\nu_{ee} \rangle| \int d|\langle m^\nu_{ee} \rangle|'
P_{1}(|\langle m^\nu_{ee} \rangle|,N_{exp}) P_{2}(|\langle m^\nu_{ee} \rangle|')
\delta (|\langle m^\nu_{ee} \rangle|-|\langle m^\nu_{ee} \rangle|') }{\sqrt{\int
P^2_{1}(|\langle m^\nu_{ee} \rangle|,N_{exp}) d|\langle m^\nu_{ee} \rangle|}
\sqrt{\int P^2_{2}(|\langle m^\nu_{ee} \rangle|) d|\langle m^\nu_{ee} \rangle|}}
\label{pp}
\end{equation}

In Table~\ref{tab:q1} we show the number of experiments needed to
reject, at 90\%, 95\%, 99\%,  and 99.73 \%CL (3 $\sigma$), the
hypothesis that neutrinos are Majorana particles. In all cases, we
find that many experiments are required in order to establish that
neutrinos are not Majorana particles. However, there is a wide
dispersion, from 13 experiments to 250 experiments at $3\sigma$, in
the number of experiments that is required depending on the
assumed pdf of the nuclear factor $F_N$.\footnote{We checked our
results by comparing with a conservative
 case. We assumed that the lightest neutrino mass is zero, neglected $\theta_{13}$ and
 the solar mass splitting,  and chose the Majorana phase to be
$\pi$. In this special case, the neutrinoless double beta mass
element has a lower limit~\cite{mpg,Bilenky:1996cb,recent1,recent2,recent3,recent4}. We can compute
straightforwardly the probability that
 the mass matrix element derived form negative searches is higher than
 the lower bound at
 a given confidence level. As expected, this calculation indicates more
 experiments are required than we found are necessary using the full
 probability distributions.  For
example, in the case ``actual, lin'' the calculation using the
lower bound gives 14 (550) required experiments at 90\%CL
($3\sigma$).}

For the bottom line table, Table~\ref{tab:summary}, we adopt the
most conservative case in which the assumed probability
distribution function for $F_N$ is given by the actual calculated
distribution of $F_N$ values. In any event,
Table~\ref{tab:summary} shows that a few next generation
neutrinoless double beta decay experiments will not be able to
answer the question of whether neutrinos are Dirac or Majorana
particles unless neutrinoless double beta decay is actually
observed\footnote{We checked that our results do not depend very
sensitively upon the assumption that equal decades in the lightest
mass $m$ are equally probable. We made instead the extreme
assumption that the pdf of the mass $m$ is equally distributed on
a linear scale with $0 < m < 2.3$ eV. This optimistic assumption
presumes that there is a 50\% chance that the lowest mass lies
between 1.15 eV and 2.3 eV. Nevertheless,the required number of
experiments at $3\sigma$ is 81.}.

What could be the effect of future improvement in the neutrino
 oscillation parameters, such as $\sin^2 \theta_{12}$ from a SNO study
with neutral current detectors (NCD) \cite{NCD},
and $\Delta m^2_{23}$ from NuMI/MINOS \cite{MINOS} and
T2K experiments \cite{T2K} ?.  In curves
labeled ``Eq. 4'' in
Fig.~\ref{fig:question1}, the two peaks correspond to the maximally
constructive (right) and destructive (left) case in
Eq.~(\ref{eq_mee_ih}), connected by a plateau due to the randomly
assigned complex phases.  On the other hand, the tails above and below
the peaks are mostly due to the uncertainties in $\sin^2 \theta_{12}$
and $\Delta m^2_{23}$.  It is clear that the improvements in
measurements cannot change the situation qualitatively.  We have performed
the same analysis with twice as accurate measurements or with no errors at
all, and found that the numbers in Table~\ref{tab:q1} cannot change more
than 40\%. For example, if we assume that all of the neutrino oscillation
parameters are known with infinite precision, the required number of
experiments to obtain a $3\sigma$ result is reduced from 230 to 156.

\begin{table}[!t]
\caption{No neutrinoless double beta decay plus inverted
hierarchy. The table gives the number of neutrinoless double beta
decay experiments with sensitivity to $|\langle m^\nu_{ee}
\rangle|$ equal to what is projected for the Majorana
\protect\cite{majorana} experiment that are required to show that
neutrinos are not Majorana particles  at 90\%, 95\%, 99\% and
99.73 \%CL  if an inverted hierarchy is correct (see
Fig.~\ref{fig:hierarchy}). We consider different probability
distributions of the nuclear factor, $F_N$: gaussian, constant, or
the actual distribution of 20 different calculations (see
Fig.~\ref{fig:fn}); either using linear (lin) or logarithmic (log)
scales. \label{tab:q1}}
\begin{tabular}{ccccc}
$F_N$ pdf &  $N_{\rm exp}$ at 90 \%CL&$N_{\rm exp}$ at 95
\%CL&$N_{\rm exp}$ at 99 \%CL
&$N_{\rm exp}$ at 99.73 \%CL\\
\noalign{\smallskip}\hline
\noalign{\medskip}actual, lin & 11 & 21 & 81 & 230\\
\noalign{\medskip}actual, log  & 9 & 17 & 61 & 141\\
\noalign{\smallskip} gaussian, lin  & 3 & 4 & 8 & 13 \\
\noalign{\medskip}gaussian, log  & 16 & 23 & 50 & 83 \\
\noalign{\medskip}constant, lin  & 4 & 7 & 21 & 45\\
\noalign{\medskip}constant, log  & 24 & 40 & 95 & 156\\
\end{tabular}
\end{table}

\section{Are neutrinos Dirac particles?  No neutrinoless double beta decay but neutrino mass measured}
\label{sec:question2}

\begin{table}[!t]
\caption{No neutrinoless double beta decay plus measured neutrino
mass in ordinary beta decay. What do we need to know to conclude
in this case that neutrinos are Dirac particles? The format of
the table is similar to
Table~\ref{tab:q1} except that for Table~\ref{tab:q2}  we assume
that a neutrino mass, $m$, has been detected in ordinary beta
decay. The table gives results for two hypothesized cases, $m=0.35
\pm 0.07$ eV \protect\cite{KATRIN} and $m=0.30 \pm 0.10$ eV
\protect\cite{KATRIN}.  If the required number of experiments
depends upon whether the neutrino mass hierarchy is normal or
inverted, then the result for the normal hierarchy is written in
parentheses. The case $m = 1.0 \pm 0.05$ eV is discussed in the
text of Sec.~\ref{sec:question2}.\label{tab:q2}}
\begin{tabular}{ccccc}
$F_N$ pdf &  $N_{\rm exp}$ at 90 \%CL&$N_{\rm exp}$ at 95
\%CL&$N_{\rm exp}$ at 99 \%CL
&$N_{\rm exp}$ at 99.73 \%CL\\
\noalign{\smallskip}\hline
\noalign{\medskip} $m=0.35 \pm 0.07$ eV &&&&\\
\noalign{\smallskip}\hline
\noalign{\medskip}actual, lin & 1 & 1 & 2 & 5(6)\\
\noalign{\medskip}actual, log & 1 & 1 & 2 & 3(4)\\
\noalign{\smallskip} gaussian, lin & 1 & 1 & 1 & 1\\
\noalign{\medskip}gaussian, log & 1 & 1 & 2 & 2\\
\noalign{\medskip}constant, lin  & 1 & 1 & 1 & 1\\
\noalign{\medskip}constant, log & 1 & 1 & 2 & 4\\
\noalign{\smallskip}\hline
\noalign{\medskip} m=0.3 $\pm$ 0.1 eV &&&&\\
\noalign{\smallskip}\hline
\noalign{\medskip}actual, lin & 1 & 2 & 5(14) & 15($\infty$)\\
\noalign{\medskip}actual, log & 1 & 1(2) & 4(10) & 9($\infty$)\\
\noalign{\smallskip} gaussian, lin  & 1 & 1 & 1(3) & 2($\infty$)\\
\noalign{\medskip}gaussian, log & 1 & 2 & 5(13) & 10($\infty$)\\
\noalign{\medskip}constant, lin & 1 & 1 & 2(4) & 4($\infty$)\\
\noalign{\medskip}constant, log & 1(2) & 2(3) & 7(19) & 16($\infty$)\\
\end{tabular}
\end{table}

In this section, we make two assumptions.
\begin{enumerate}
\item Next generation experiments \cite{Gratta} do not observe neutrinoless double beta decay.
\item Next generation beta decay experiments \cite{KATRIN} observe the neutrino mass
scale.
\end{enumerate}
The first assumption is identical to our first assumption in
Sec.~\ref{sec:question1}. The second assumption assumes that an
experiment with the expected sensitivity of the  KATRIN
experiment~\cite{KATRIN} will successfully identify a spectral
distortion of the tritium beta decay energy spectrum that is due
to a finite neutrino mass.

Given the measurement of a neutrino mass in the KATRIN experiment,
how many double beta experiments would we need to establish that
neutrinos are Dirac particles at a given CL?

We will consider three cases. First, $m = 1$ eV, which is chosen
because this value is close to the present upper bound for a neutrino
mass in ordinary beta decay \cite{mainz,troitsk}. Second, $m = 0.35$
eV, which is chosen because this is the smallest mass that could be
discovered at $5 \sigma$ in next generation experiments that perform
with the sensitivity of the KATRIN experiment~\cite{KATRIN}. Third,
$m= 0.30$ eV, which is chosen because it is the smallest mass that
could be discovered at $3\sigma$ in a next generation experiment with
the expected KATRIN sensitivity~\cite{KATRIN}.  We assume that $m$ is
normally distributed with a mean value of $1.0$ ($0.35$) [$0.3$] eV
and standard deviation $0.05$ ($0.07$) ($0.10$) eV for the three cases
listed in the order given above.\footnote{It is possible that $m^2$ is
  normally distributed at the same significance rather than $m$.  Then
  $m=0.3 \pm 0.1$~eV is replaced by $m^2 = 0.09 \pm 0.03$~eV$^2$,
  and the effective error in $m$ is reduced.  In this case, we find
 that the number of experiments required to conclude neutrinos
 are Dirac particles is slightly larger (differ at most in 1(2)
experiment(s) at  99(99.73) \%CL) than in the case
$m=0.35 \pm 0.07$~eV shown in
  Table~\ref{tab:q2} .  On the other hand, it remains true that the
sensitivity to the Majorana character of neutrinos quickly
runs out of steam below 0.3~eV.}

The last two cases, which are given as examples in the Majorana
proposal~\cite{majorana},  are separated by only $0.05$ eV.
However, as we shall see in the discussion below, this small
difference in mass makes a large difference in the number of
required experiments. The essential reason for this large
difference is that if experiment shows that $m = 0.35 \pm 0.07$
eV, then we know that $m$ is well separated from zero mass at
$3\sigma$. However, if $m = 0.30 \pm 0.10$ eV, then at $3\sigma$
the lightest mass could be zero.

We are now in a position to compute the required number of
neutrinoless double beta decay experiments. Compared to the
analysis done in Sec.~\ref{sec:question1}, we need only modify our
analysis of Eq.~(\ref{eq_mee_ih}), replacing the pdf assumed in
Sec.~\ref{sec:question1} for $m$ by the corresponding gaussian
distribution that represents one of the three cases listed above
for a next generation beta decay experiment. Moreover, in this
section we do calculations for both neutrino mass hierarchies,
normal and inverted [given by Eq.~(\ref{eq_mee_ih})]. For a normal
hierarchy, the neutrino mass element can be written as
\begin{equation}
|\langle m^\nu_{ee} \rangle|_{NH} = \left| \cos^2\theta_{13} ( m
\cos^2\theta_\odot + \sqrt{m^2+\Delta m^2_\odot}
\sin^2\theta_{\odot}  ~e^{i\phi_1} ) + \sqrt{m^2+\Delta m^2_{atm}
+\Delta m^2_\odot} \sin^2\theta_{13}  ~e^{i\phi_2}\right| \,
.\label{eq_mee_nh}
\end{equation}

For the first case, $m = (1.0 \pm 0.05)$ eV, one experiment is
sufficient to prove that neutrinos are Dirac particles at more
than 3$\sigma$ ($P > 99.77~\%$).

 Table~\ref{tab:q2} presents for the second and third cases the results
for different assumptions
about the pdf of $F_N$. For the second case, $m = 0.35 \pm 0.07$
eV, one experiment is sufficient to prove that neutrinos are Dirac
particles at 95 \% CL but 6 experiments are required to prove that
neutrinos are not Dirac particles at $3\sigma$. For the third
case, $m = (0.30 \pm 0.10)$ eV, and assuming an inverted
hierarchy, 2 experiments are sufficient to prove neutrinos are
Dirac particles at 90 \% CL but 16 experiments are required to
prove that neutrinos are not Majorana particles at $3\sigma$.

The differences between hierarchies are small in the first and
second cases listed above because the mass scale $m$ is assumed
large compared with the solar and atmospheric mass splittings. In
these two cases, the neutrino masses are essentially degenerate
(imagine Fig.~\ref{fig:hierarchy} for the case in which $m$ is
much larger than either the solar or the atmospheric mass
splitting).

For the third case, $m = (0.030 \pm 0.10) $ eV, there are large
quantitative differences  between the normal hierarchy and the
inverted hierarchy at higher CL.
 The reason for the difference in behavior can be seen visually in
 Fig.~\ref{fig:mee_m1}. The mass matrix element $|\langle
m^\nu_{ee} \rangle|$ in a normal hierarchy can be greatly reduced
because of cancellations, while $m$ is bounded from below in an
inverted mass hierarchy. For a normal hierarchy and small values
of $m$, $|\langle m^\nu_{ee} \rangle|$ could be extremely small,
orders of magnitude below the expected level of sensitivity of
next generation double beta decay experiments. Therefore, the
corresponding entries in Table~\ref{tab:q2} require at $3\sigma$
CL an infinite number of next generation double beta decay
experiments to distinguish between Dirac and Majorana particles.

\section{What is the total mass in neutrinos?  Neutrinoless double beta decay detected}
\label{sec:question3}

\begin{figure}[!ht]
  \centerline{\includegraphics[width=4.5in]{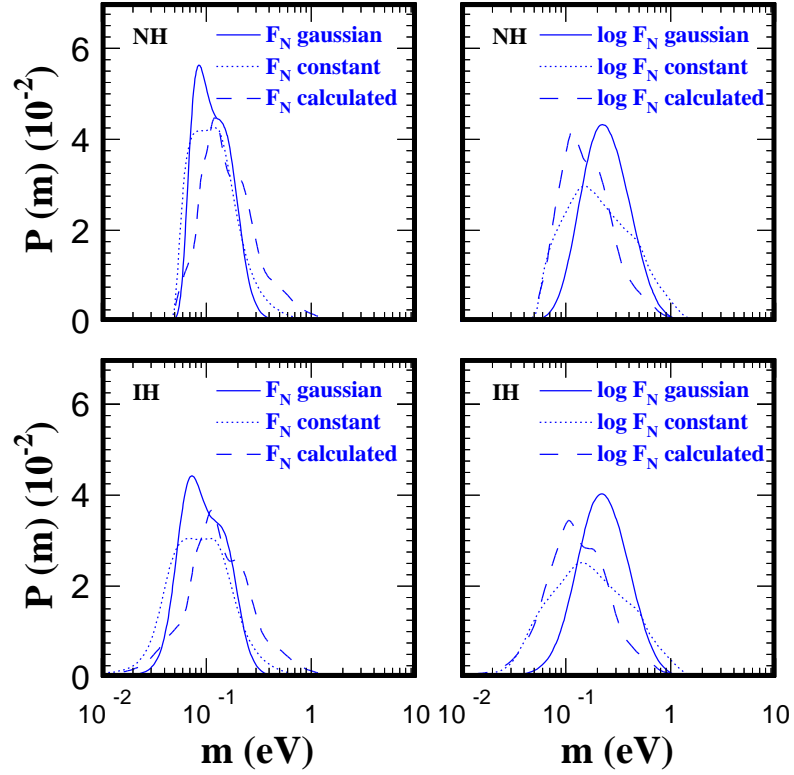}}
  \caption{Probability distributions of the lightest mass eigenstate
    m in a future generation neutrinoless double beta decay
    experiments. For illustration, we assumed a measured half life
    $T_{\frac{1}{2}}(^{76}{\rm Ge})=3.2 \pm 0.5 \times 10^{26}$ yr. We
    obtained the distributions with the aid of Eq.~(\ref{pp2}) (see
    text for details). For the left hand panels, we assume that the
    probability distribution function for the nuclear factor $F_N$
    that appears in Eq.~(\ref{eq:matrixelementdefn}) satisfies a
    Gaussian distribution (full line), a constant distribution (dotted
    line), or the distribution of computed nuclear factor calculations
    by different theoretical groups (dashed line). The right hand
    panels correspond to assuming that $\log F_N$ follows those
    distributions. The upper panels correspond to a normal neutrino
    mass hierarchy [Eq.~(\ref{eq_mee_nh})], while the lower panels
    correspond to an inverted hierarchy [Eq.~(\ref{eq_mee_ih})]. }
\label{fig:question3}
\end{figure}

We suppose in this section that next generation
experiments\cite{Gratta} successfully detect neutrinoless double
beta with a large neutrino mass matrix element $|\langle
m^\nu_{ee} \rangle|$.

We will compute in this section the pdf for the lowest mass
eigenstate, $m$, using hypothesized results from next generation
neutrinoless double beta decay experiments. Since we already know
from existing experiments the pdf for the mass splittings, $\Delta
m^2_\odot$ and $\Delta m^2_{atm}$, we can use these data together
with the results for the lowest mass $m$ to compute the cumulative
pdf for the total mass in neutrinos.

We will use $^{76}$Ge as an illustrative case. The
Heidelberg-Moscow experiment~\cite{Heidelberg-Moscow} provides a
lower limit on the half life (we remind the reader that there is a
claim of 4$\sigma$ detection in Ref.~\cite{Klapdor-Kleingrothaus:2004ge}),
\begin{equation}
T_{1/2} ~>~ 1.9 (3.1) \times 10^{25} \, {\rm yr}
\label{eq:heidelbergmoscow}
\end{equation}
at 90\% CL (68\% CL) .

We consider three feasible cases with positive neutrinoless double
beta detection: a) $T_{1/2} = (3.2 \pm 0.2) \times 10^{25}$ yr, b)
$T_{1/2} = (1.\pm 0.1) \times 10^{26}$ yr, and c) $T_{1/2} = (3.2
\pm 0.5) \times 10^{26}$ yr, corresponding to 373, 118, and 37.3
events expected in the parameter region of interest in the
Majorana experiment \cite{majorana}. The expected background in a
deep underground experiment is 5.5 events, although background
could be different by a factor of two. Systematic errors are
expected to be a few percent and to be dominated by energy
resolution, the segmentation cut, and the pulse shape
discrimination acceptance. Our results are not significantly
affected by including systematic errors of a few percent because
of the dominant contribution of the uncertainty in the nuclear
factor $F_N$.

We computed numerically the pdf of the neutrino mass element
$|\langle m^\nu_{ee} \rangle|$ given by
Eq.~(\ref{eq:matrixelementdefn}) for all three values of
$T_{\frac{1}{2}} $ listed above and for all six possible
distributions of the nuclear factor $F_N$ that were discussed in
Sec.~\ref{sec:question1} and Sec.~\ref{sec:question2}. The
determination of the neutrino mass element can be used to extract
the probability distribution, $P(m)$, of the lightest mass
eigenstate by extracting with the help of Eq.~(\ref{eq_mee_ih}) and
Eq.~(\ref{eq_mee_nh}). In order to find $P(m)$, we compute

\begin{equation}
P (m)= \frac{\int d|\langle m^\nu_{ee} \rangle| \int d|\langle
m^\nu_{ee} \rangle|' P_{1}(|\langle m^\nu_{ee} \rangle|)
P_{2}(|\langle m^\nu_{ee} \rangle|',m) \delta (|\langle m^\nu_{ee}
\rangle|-|\langle m^\nu_{ee} \rangle|') }{\sqrt{\int
P^2_{1}(|\langle m^\nu_{ee} \rangle|) d|\langle m^\nu_{ee}
\rangle|} \sqrt{\int P^2_{2}(|\langle m^\nu_{ee} \rangle|,m)
d|\langle m^\nu_{ee} \rangle|}}\, , \label{pp2}
\end{equation}
where $P_{1}(|\langle m^\nu_{ee} \rangle|)$ is the pdf of the
neutrino mass element given by Eq.~(\ref{eq:matrixelementdefn}) and
$P_{2}(|\langle m^\nu_{ee} \rangle|,m)$ is the pdf of the neutrino
mass element given by Eq.~(\ref{eq_mee_nh}) or Eq.~(\ref{eq_mee_ih}),
respectively, for a normal or an inverted neutrino mass hierarchy.

\begin{figure}[!ht]
  \centerline{\includegraphics[width=4.5in]{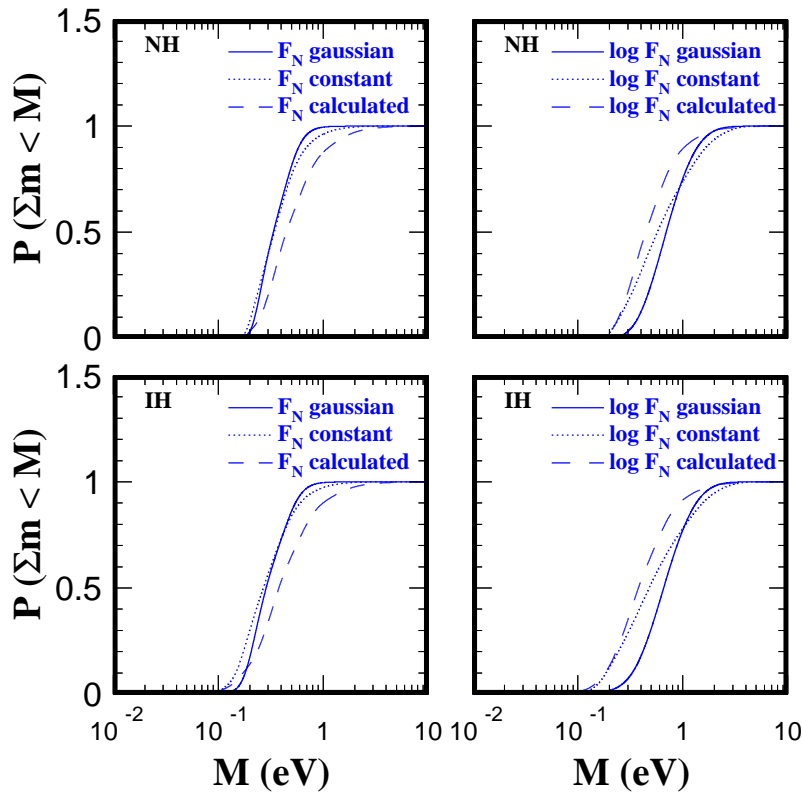}}
  \caption{The cumulative probability that the total mass in
    neutrinos is less than $M$. For illustration, we assumed a measured
    neutrinoless double beta decay half life
    $T_{\frac{1}{2}}(^{76}{\rm Ge})=3.2 \pm 0.5 \times 10^{26}$ yr. We
    calculated the cumulative probability by integrating
    Eq.~(\ref{pp2}) (see text for details). The organization of the
    panels and the notation are the same as for
    Fig.~\ref{fig:question3}.} \label{fig:question3tot}
\end{figure}

Figure~\ref{fig:question3} shows the computed pdfs of the lightest
mass eigenstate in the case of normal and inverted neutrino mass
hierarchies for different assumptions regarding the pdf of the
nuclear factor $F_N$. For illustrative purposes, we assumed in
making the figure that $T_{1/2} = (3.2 \pm 0.5) \times 10^{26}$ yr
(case c above). The two other lifetimes considered above, case a
and case b, result in pdfs with very similar shapes but shifted
relative to Fig.~\ref{fig:question3} to larger values of $m$, the
lightest neutrino mass.

We can also extract from our analysis the allowed ranges of the
total mass in neutrinos at a given CL. Table \ref{tab:question3}
presents the allowed ranges for the total mass in neutrinos for
different assumptions regarding the pdfs of the nuclear factor
$F_N$ and for the three values of the half-life for neutrinoless
double beta decay assumed above (cases a-c).

Figure~\ref{fig:question3tot} shows the cumulative probabilities
for the total mass in neutrinos $M$ ($M=m_1+m_2+m_3$).  The results
are illustrated for different regarding the pdf of $F_N$ and for
both normal and inverted neutrino mass hierarchies.  In
constructing Fig.~\ref{fig:question3tot}, we assumed that $T_{1/2}
= (3.2 \pm 0.5) \times 10^{26}$ yr (case c above). For the shorter
half-lives corresponding to cases a) and b) above, the cumulative
probabilities have very similar shapes but are shifted to larger
values of the total neutrino mass.

\begin{table}[!t]
\caption{Allowed ranges of the total mass in neutrinos for
different assumed measurements of the half-life of $^{76}$Ge to
neutrinoless double beta decay.  We consider different probability
distributions of the nuclear factor: gaussian, constant, or the
actual distribution of 20 different theoretical calculations;
using either a linear (lin) or a logarithmic (log) scale for
$F_N$. In general, the results are different for normal and for
inverted neutrino mass hierarchies. The results for the normal
hierarchy are written in parenthesis. \label{tab:question3}}
\begin{tabular}{ccccc}
$F_N$ pdf &  $M$ (eV) at 90 \%CL&$M$ (eV) at 95 \%CL&$M$ (eV) at 99 \%CL
&$M$ (eV) at 99.73 \%CL\\
\noalign{\smallskip}\hline
\noalign{\medskip} $T_{1/2} = (3.2 \pm 0.2) \times 10^{25}$ yr\\
\noalign{\smallskip}\hline
\noalign{\medskip}actual, lin & [0.63,4.70] ([0.65,4.73]) & [0.56,5.96] ([0.58,5.99])
& [0.48,8.58] ([0.50,8.60]) & [0.46,9.56] ([0.48,9.58])\\
\noalign{\medskip}actual, log & [0.64,4.22] ([0.66,4.24]) & [0.57,5.36] ([0.59,5.39])
& [0.49,7.84] ([0.51,7.88]) & [0.46,9.41] ([0.48,9.43])\\
\noalign{\smallskip} gaussian, lin & [0.62,2.07] ([0.63,2.09]) & [0.59,2.35] ([0.60,2.36])
& [0.54,3.09]  ([0.55,3.10]) & [0.51,3.93] ([0.52,3.94])\\
\noalign{\medskip}gaussian, log & [0.96,5.17] ([0.98,5.17]) & [0.84,6.04] ([0.85,6.04])
& [0.64,7.98] ([0.66,7.98]) & [0.53,9.24] ([0.55,9.24])\\
\noalign{\medskip}constant, lin  & [0.55,2.81] ([0.57,2.83]) & [0.52,3.61] ([0.54,3.63])
& [0.48,5.90] ([0.50,5.92]) & [0.46,7.88] ([0.48,7.90])\\
\noalign{\medskip}constant, log & [0.63,6.45] ([0.65,6.48]) & [0.57,7.74] ([0.59,7.75])
& [0.51,9.45] ([0.53,9.45]) & [0.48,9.83] ([0.50,9.83])\\
\noalign{\smallskip}\hline
\noalign{\medskip}$T_{1/2} = (1.0\pm 0.1) \times 10^{26}$ yr\\
\noalign{\smallskip}\hline
\noalign{\medskip}actual, lin & [0.34,2.76] ([0.37,2.81]) & [0.30,3.63] ([0.33,3.69])
& [0.26,6.23] ([0.29,6.30]) & [0.24,8.34] ([0.28,8.40])\\
\noalign{\medskip}actual, log & [0.34,2.46] ([0.38,2.52]) & [0.30,3.24] ([0.34,3.29])
& [0.26,6.20] ([0.30,6.29]) & [0.24,8.06] ([0.28,8.11])\\
\noalign{\smallskip} gaussian, lin  & [0.34,1.16] ([0.36,1.18]) & [0.32,1.32] ([0.34,1.34])
& [0.29,1.73] ([0.31,1.76]) & [0.28,2.21] ([0.30,2.24])\\
\noalign{\medskip}gaussian, log & [0.53,2.93] ([0.56,2.94]) & [0.46,3.43] ([0.49,3.45])
& [0.34,4.65] ([0.38,4.66]) & [0.28,5.70] ([0.32,5.71])\\
\noalign{\medskip}constant, lin & [0.29,1.56] ([0.33,1.61])& [0.28,2.01] ([0.31,2.07])
& [0.26,3.35] ([0.29,3.43]) & [0.24,4.66] ([0.28,4.73])\\
\noalign{\medskip}constant, log & [0.34,3.80] ([0.37,3.86]) & [0.31,4.68] ([0.34,4.73])
& [0.27,6.33] ([0.30,6.36]) & [0.25,7.46] ([0.29,7.47])\\
\noalign{\smallskip}\hline
\noalign{\medskip}$T_{1/2} = (3.2 \pm 0.5) \times 10^{26}$ yr\\
\noalign{\smallskip}\hline
\noalign{\medskip}actual, lin & [0.14,1.45] ([0.22,1.60])& [0.12,1.95] ([0.19,2.13])
& [0.09,3.63] ([0.17,3.93]) & [0.08,5.68] ([0.16,6.06])\\
\noalign{\medskip}actual, log & [0.15,1.29] ([0.22,1.42])& [0.12,1.72] ([0.20,1.85])
& [0.09,3.31] ([0.17,3.56]) & [0.08,4.51] ([0.16,4.67])\\
\noalign{\smallskip} gaussian, lin & [0.16,0.63] ([0.21,0.67])& [0.15,0.72] ([0.20,0.76])
& [0.12,0.95] ([0.18,1.00]) & [0.09,1.20] ([0.17,1.27])\\
\noalign{\medskip}gaussian, log & [0.27,1.62] ([0.32,1.66])& [0.22,2.61] ([0.28,1.94])
& [0.13,2.61] ([0.22,2.64]) & [0.10,3.21] ([0.19,3.25])\\
\noalign{\medskip}constant, lin  & [0.13,0.79] ([0.19,0.91]) & [0.11,1.02] ([0.18,1.16])
& [0.09,1.74] ([0.17,1.93]) & [0.08,2.46] ([0.16,2.67])\\
\noalign{\medskip}constant, log & [0.16,2.03] ([0.22,2.16])& [0.14,2.53] ([0.20,2.65])
& [0.10,3.51] ([0.18,3.60]) & [0.09,4.18] ([0.17,4.26])\\
\end{tabular}
\end{table}

\section{Normal or Inverted Mass Hierarchy?  Neutrinoless double beta decay detected but neutrino mass not measured}
\label{sec:question4}

 We make two assumptions in this section.
\begin{enumerate}
\item Next generation experiments \cite{Gratta} will observe
neutrinoless double beta decay. \item Next generation ordinary
beta decay experiments~\cite{KATRIN} will not detect the neutrino
mass scale.
\end{enumerate}
These assumptions are the opposite of what we postulated in
Sec.~\ref{sec:question2}.

In this section, we answer the following question. Given the
detection of neutrinoless double beta decay and the non detection
of a neutrino mass in normal beta decay,  can we determine if the
neutrino mass hierarchy is normal or inverted?

In order to answer this question, we computed the $\chi^2$
distribution as a function of the different neutrino variables,
including neutrino oscillation data where available for $\Delta
m^2_\odot$, $\Delta m^2_{atm}$, $\theta_{\odot}$, $\theta_{13}$,
the lightest mass m, the Majorana phases $\phi_1$ and $\phi_2$,
and the neutrinoless mass $|\langle m^\nu_{ee} \rangle|$. For an
inverted (normal) neutrino mass hierarchy, we imposed
Eq.~(\ref{eq_mee_ih}) [Eq.~(\ref{eq_mee_nh})].
 We then marginalized over all variables except $m$ and $|\langle m^\nu_{ee} \rangle|)$.

Figure~\ref{fig:mee_m1} shows the allowed regions in the $|\langle
m^\nu_{ee} \rangle|)$-$m$ plane at 90\% CL for the inverted and
normal hierarchy (full regions labelled IH and NH).

Just as we did in Sec.~\ref{sec:question3}, we consider three
cases of positive neutrinoless double beta detection with a
$^{76}$Ge half life : a) $T_{\frac{1}{2}} = (3.2 \pm 0.2) \times
10^{25}$ yr, b) $T_{\frac{1}{2}} = (1.\pm 0.1) \times 10^{26}$ yr,
and c) $T_{\frac{1}{2}} = (3.2 \pm 0.5) \times 10^{26}$ yr.

Normal and inverted neutrino mass hierarchies can not be
distinguished solely by a positive signal in a neutrinoless double
beta decay next generation experiment. This is illustrated by the
dashed lines in Fig.~\ref{fig:mee_m1} corresponding to cases
a)-c). All other things being equal, a relatively large value for
$|\langle m^\nu_{ee} \rangle|)$ favors an inverted hierarchy.
However, for any experimentally accessible value of $|\langle
m^\nu_{ee} \rangle|)$ that is inferred from neutrinoless double
beta decay, one can always postulate a sufficient large value of
the lowest neutrino mass, $m$, that would account for the measured
decay rate with mass-degenerate neutrinos.

In order to distinguish between a normal and an inverted neutrino
mass hierarchy, we must somehow know that the lowest mass
eigenstate $m$ is very small (less than $0.01$ eV).  If we had a
private communication showing that the lowest neutrino mass were
zero, then we could distinguish between a normal and an inverted
mass hierarchy. We find from detailed calculations that all three
of the hypothetically successful measurements of a double beta
decay lifetime [cases a), b), and c) above] would, if $m=0$,
exclude a normal hierarchy independent of the pdf of the nuclear
factor $F_N$.

\section{Summary and Discussion} \label{sec:summary}

Next generation neutrinoless double beta decay experiments offer
the promise of a fundamental discovery, namely, that neutrinos are
their own anti-particles. No other feasible experimental
technique could establish this profound result. If a single
experiment conclusively detects zero neutrino double beta decay,
then weak interaction theory will be both profoundly simplified
and greatly clarified.

Even if neutrinoless double beta decay is not observed in next
generation experiments, we may still be able to conclusively
determine the particle/anti-particle nature of neutrinos. If an
ordinary beta-decay experiment detects a neutrino mass near 1 eV,
then we will be able to conclude in this case that neutrinos are
Dirac not Majorana particles.

In all other cases, the situation will be much less favorable, as
can be seen readily from the summary given in
Table~\ref{tab:summary}. If ordinary beta-decay reveals a neutrino
mass scale of less than 0.3 eV, then we will not be able to
conclude that neutrinos are Dirac particles from the
non-observation of neutrinoless double beta decay
in currently envisioned experiments. The
particle/anti-particle nature of neutrinos will remain ambiguous.

The observation of neutrinoless double beta decay will determine a
large allowed range of the total mass in the form of neutrinos, a
range that permits an uncertainty in the total mass of between one
and two orders of magnitude.  This range translates into a total
cosmic neutrino mass density ~ (cf. \cite{WMAP}) $\Omega_{\nu}
=0.009-0.20$, $\Omega_{\nu}=0.005-0.17$, or $\Omega_{\nu}
=0.0016-0.12$ at 3$\sigma$ for the three assumed lifetimes listed
in Table~\ref{tab:summary} and discussed in
Sec.~\ref{sec:question3}.

Finally, we note that we will not be able to decide whether the
neutrino mass hierarchy is normal or inverted even if neutrinoless
double beta decay is detected. In order to decide this important
question, information from other types of experiments like long
baseline oscillation studies will be necessary.

\acknowledgments JNB wishes to thank S. R. Elliott and A. Giuliani for
their excellent review talks on neutrinoless double beta-decay at
TAUP03, which stimulated this investigation. JNB and
CPG acknowledge support from NSF Grant No. PHY-0070928. HM was
supported by the Institute for Advanced Study, funds for Natural
Sciences.  His work was also supported in part by the DOE under
contracts DE-AC03-76SF00098 and in part by NSF Grant No. PHY-0098840.

\appendix
\section{Upper and lower bounds connected with neutrinoless double beta decay}

In this appendix, we derive an upper bound,
Sec.~\ref{subsec:appendixupperbound}, and a lower bound,
Sec.~\ref{subsec:appendixlowerbound},  on $|\langle m^\nu_{ee}
\rangle|$.  We assume that neutrinoless double beta decay is not
observed in next generation experiments and that the neutrino mass
hierarchy is inverted. In Sec.~\ref{subsec:approximate}, we obtain
approximate results for the number of experiments that are
required to show that neutrinos are Dirac particles using the
inequalities derived in Sec.~\ref{subsec:appendixupperbound} and
Sec.~\ref{subsec:appendixlowerbound}.

\subsection{An upper bound on $|\langle m^\nu_{ee} \rangle|$}
\label{subsec:appendixupperbound}
If a neutrinoless double beta decay experiment does not detect any
events above the expected background, then the half-life satisfies
\begin{equation}
T_{1/2} ~\ge~ \frac{\Delta t \log2}{-\log\alpha} N_X \epsilon \, ,
\label{eq:tonehalfdefn}
\end{equation}
where $\Delta t$ is the period of data taking, $N_X$ is the total number
 of active nuclei $X$ and $\epsilon$ is the efficiency of event capture
after cuts to reduce background. The quantity $\alpha = 1-\%
{\rm CL}/100$ is a given significance level. For definiteness, we will
use the expectations for the Majorana experiment~\cite{majorana}
to determine a reference sensitivity $s$ to $T_{\frac{1}{2}}F_N$
[see Eq.~(\ref{eq:matrixelementdefn})] in next generation
neutrinoless double beta experiments \cite{Gratta}. The Majorana
collaboration \cite{majorana} is planning to use a  500 kg Ge
(86\% $^{76}$Ge) detector, $\Delta t =$ 5 yr, and
$\epsilon=60$\%. With these values of the parameters,
Eq.~(\ref{eq:tonehalfdefn}) becomes
\begin{equation}
T_{1/2} (Ge) ~\ge~ \frac{7.13\times10^{27}}{-\log\alpha} {\rm ~
yr} \,. \label{eq:tonehalfnumbers}
\end{equation}

Different nuclear structure parameter $F_N$ calculations of the
transition in the case of $^{76}$Ge, about 20, expand over a range
(that we will consider as a 3$\sigma$ range determination) of
\begin{equation}
F_N ~=~ (1.455 \pm 1.425)\times10^{-13} {\rm yr}^{-1} \,
.\label{eq:fnnumbers}
\end{equation}
The distribution of calculated values of $F_N$ is shown in
Fig.~\ref{fig:fn}.

Inserting Eqs.(\ref{eq:tonehalfnumbers}) and ~(\ref{eq:fnnumbers}) in
Eq.~(\ref{eq:matrixelementdefn}), we find
\begin{equation}
|\langle m^\nu_{ee} \rangle| ~\le~ 1.913\times10^{-2}
\sqrt{\frac{-\log\alpha} {1.455 + 0.475 n(\alpha)}} ~~{\rm eV}\, ,
\end{equation}
where $n(\alpha)$ is the number of standard deviations at a given CL, with an
asymptotic expansion
\begin{equation}
n(\alpha) = \sqrt{\log\left(\frac{2}{\pi \alpha^2}\right) -
\log\left(\log\left(\frac{2}{\pi \alpha^2}\right)\right)}\, .
\end{equation}

 For $N$ neutrinoless double beta decay experiments, $N_{exp}$, with
 sensitivity to $|\langle m^\nu_{ee} \rangle|$ of $s'$, we have
\begin{equation}
|\langle m^\nu_{ee} \rangle| ~\le~ 1.913\times10^{-2}
\sqrt{\frac{s}{N_{exp} s'}} \sqrt{\frac{-\log\alpha}{1.455 + 0.475
n(\alpha)}} ~~{\rm eV} \, . \label{mee1}
\end{equation}

\subsection{Inverted hierarchy : A lower bound}
\label{subsec:appendixlowerbound}

 If the neutrino mass hierarchy
is inverted, the neutrino mass element $|\langle m^\nu_{ee}
\rangle|$ can be related to neutrino parameters determined in
oscillation experiments \cite{mpg,Bilenky:1996cb,recent1,recent2,recent3,recent4}
(see Fig.~\ref{fig:mee_m1} for illustration) by the relation
\begin{equation}
|\langle m^\nu_{ee} \rangle| ~\ge~ \sqrt{\Delta m_{atm}^2
\cos{2\theta_{\odot}}} \sim
\left(\frac{-\log{(1-\alpha)}}{3\times10^{17}}\right)^{1/10}\, .
\label{eq:meelowerbound}
\end{equation}
The  fitting-function with the exponent of 1/10 that appears in
Eq.~(\ref{eq:meelowerbound}) reproduces well the results obtained in
the analysis of solar and atmospheric data~\cite{mpg}. The
fitting-function in Eq.~(\ref{eq:meelowerbound}) deviates from the
numerical results by less than 1\% in the range [20, 99.9] \% CL.

\subsection{Approximate answer to the question posed in Sec.~\ref{sec:question1}}
\label{subsec:approximate}

Eq.~(\ref{mee1}) and Eq.~(\ref{eq:meelowerbound}) can be used to
determine approximately the number of experiments, $N_{exp}$, with
the expected sensitivity of the Majorana experiment that are
required to show that neutrinos are Dirac particles if
neutrinoless double beta-decay is not observed and if the neutrino
mass hierarchy is inverted.  This question was answered by a brute
force method in Sec.~\ref{sec:question1}. By requiring that there
be no intersection of the inequalities Eq.~(\ref{mee1}) and
Eq.~(\ref{eq:meelowerbound}), we calculate that the number of
experiments required  is $N_{exp} \ge$ 7, 12, 57, and 645 at 90,
95, 99 and 99.73 \% CL, respectively. The approximate results
obtained here are in good agreement with the more accurate results
obtained in Sec.~\ref{sec:question1} and listed in
Table~\ref{tab:q1}.

\section{Sensitivity of proposed neutrinoless double beta decay experiments}

\begin{table}[!t]
\caption{Proposed or suggested neutrinoless double beta decay
experiments. The half-life sensitivities  are estimated by the
collaborations (with assumptions about backgrounds that have not
yet been established experimentally) and scaled for 5 years of
data taking (updated form Ref.~\protect\cite{ev}). The last row of
Table~\ref{tab:0nubbfut} presents the calculated sensitivities to
the neutrinoless double beta-decay mass matrix element, $|\langle
m^\nu_{ee} \rangle|$. To compute results in last row, we used the
published distributions of calculated nuclear matrix
elements~\protect\cite{calc} on a logarithmic scale. If less than
three independent calculations of $F_N$ have been published for a
given nucleus, no estimate was computed for the sensitivity for
that nucleus to the neutrino mass matrix element.}
\label{tab:0nubbfut}
\begin{center}
\begin{tabular}{cccc}
&&  Sensitivity to &  Sensitivity to         \\
Experiment     &    Source  & $T_{\frac{1}{2}}$ (yr) at 90\% CL
& $|\langle m^\nu_{ee}\rangle|$ (eV) at 90\% CL   \\ \hline
CANDLES\cite{KIS01}        &$^{48}$Ca          & $1 \times 10^{26}$    & 0.248 \\
Majorana\cite{majorana}    &$^{76}$Ge          & $3 \times 10^{27}$    & 0.054 \\
GEM\cite{ZDE01}            &$^{76}$Ge          & $7 \times 10^{27}$    & 0.034 \\
GENIUS\cite{KLA01b}        &$^{76}$Ge          & $1 \times 10^{28}$    & 0.028 \\
NEMO 3\cite{NEMO3}         &$^{100}$Mo         & $4 \times 10^{24}$    & 0.646 \\
MOON\cite{EJI00}           &$^{100}$Mo         & $1 \times 10^{27}$    & 0.041 \\
CAMEO\cite{BEL01}          &$^{116}$Cd         & $1 \times 10^{27}$    & 0.057 \\
COBRA\cite{ZUB01}          &$^{130}$Te         & $1 \times 10^{24}$    & 1.260 \\
CUORICINO\cite{AVI}        &$^{130}$Te         & $1.5 \times 10^{25}$  & 0.336 \\
CUORE\cite{AVI}            &$^{130}$Te         & $7 \times 10^{26}$    & 0.049 \\
XMASS\cite{XMASS}          &$^{136}$Xe         & $3 \times 10^{26}$    & 0.134 \\
Xe\cite{CAC01}             &$^{136}$Xe         & $5 \times 10^{26}$    & 0.104 \\
EXO\cite{Gratta}           &$^{136}$Xe         & $1 \times 10^{28}$    & 0.023 \\
DCBA\cite{ISH00}           &$^{150}$Nd         & $2 \times 10^{25}$    & 0.498\\
GSO\cite{DAN01,WANGS01}    &$^{160}$Gd         & $2 \times 10^{26}$    & --- \\
\\\hline

\end{tabular}

\end{center}
\end{table}

Several next generation neutrinoless double beta experiments have
been proposed. Table~\ref{tab:0nubbfut}
 lists a representative sample of different nuclei for
 which neutrinoless double beta-decay experiments have been
 proposed
(updated from Ref.~\cite{ev}). The claimed sensitivity is shown in
the third column of Table~\ref{tab:0nubbfut}, quantified by the
half-life limit at 90\% CL in the case of negative searches. These
limits have been evaluated using assumptions on background rates
that have not yet been demonstrated experimentally and are scaled
for 5 years of data taking. The comparison between different
experiments should be made taking  these considerations into
account. The last column presents the sensitivity to the
neutrinoless mass element, $|\langle m^\nu_{ee} \rangle|$. We used
the distributions of calculated nuclear matrix element given in
Ref.~\cite{calc}. As far as we know, this has not been done in
previous publications.
 In the literature, the translation from half-life to mass is usually
 made for a particular assumed nuclear matrix element factor.
 We think it may be useful to describe, as in Table~\ref{tab:0nubbfut},
 the obtainable limits on the neutrinoless mass element by using
the   distribution of the calculated nuclear matrix elements that
has been published for each nucleus.

Most nuclei have not yet been studied as widely as $^{76}$Ge. We
have determined a limit on the neutrinoless mass matrix element if
there are three or more published nuclear matrix element
calculations. We did not compute a limit for $^{160}$Gd, for which  we
found only 2 published calculations \cite{Tretyak:1995rj,calc}.

\end{document}